# Charge-Density-Wave Devices Printed with the Ink of Chemically Exfoliated 1T-TaS$_2$ Fillers

Saba Baraghani,[1,2] Zahra Barani,[1] Yassamin Ghafouri,[3] Amirmahdi Mohammadzadeh,[1] Sriharsha Sudhindra,[1] Tina T. Salguero,[3] Fariborz Kargar,[1,2,*] and Alexander A. Balandin[1,2,*]

[1]Nano-Device Laboratory and Phonon Optimized Engineered Materials Center, Department of Electrical and Computer Engineering, University of California, Riverside, California 92521 USA

[2]Department of Chemical and Environmental Engineering, University of California, Riverside, California 92521 USA

[3]Department of Chemistry, University of Georgia, Athens, Georgia 30602 USA

---

[*] Corresponding authors: fkargar@ece.ucr.edu ; balandin@ece.ucr.edu ; web-site: http://balandingroup.ucr.edu/

## Abstract

We report on the preparation of inks containing fillers derived from quasi-two-dimensional charge-density-wave materials, their application for inkjet printing, and the evaluation of their electronic properties in printed thin film form. The inks were prepared by liquid-phase exfoliation of CVT-grown 1T-TaS$_2$ crystals to produce fillers with nm-scale thickness and μm-scale lateral dimensions. Exfoliated 1T-TaS$_2$ was dispersed in a mixture of isopropyl alcohol and ethylene glycol to allow fine-tuning of their thermo-physical properties for inkjet printing. The temperature-dependent electrical and current fluctuation measurements of printed thin films demonstrated that the charge-density-wave properties of 1T-TaS$_2$ are preserved after processing. The functionality of the printed thin-film devices can be defined by the nearly-commensurate to commensurate charge-density-wave phase transition of individual exfoliated 1T-TaS$_2$ fillers rather than by electron-hopping transport between them. These results provide pathways for the development of printed electronics with diverse functionality achieved by the incorporation of quasi-two-dimensional van der Waals quantum materials.

## Introduction

Solution-processed two-dimensional (2D) van der Waals materials offer a scalable route toward the next generation of flexible printed electronics. For example, the inks of 2D materials are essential for wearable and skin electronics based on mechanically soft and stretchable materials compatible with such integration.[1–6] Inkjet printing is one of the most promising approaches for large-area fabrication of flexible electronics. Printing of a wide range of electronic components, including metal wiring and interconnects, field-effect transistors, capacitors, photovoltaic devices, and light-emitting diodes has been demonstrated.[7–11] Because inkjet printing utilizes a limited number of process steps, it is well suited for mass production.

However, there is one serious issue with printed electronics based on 2D material inks: the valuable intrinsic properties of semiconducting materials are compromised by electron hopping mechanisms. That is, one can readily print a metallic conductor or an electrical insulator. A percolated network of graphene or metallic carbon nanotubes can conduct the electricity efficiently, offering a rather low sheet resistance.[12–14] A printed layer of boron nitride ink can serve as an efficient electrically insulating barrier, serving, for instance, as a gate dielectric. However, a printed layer of an ink with semiconducting fillers never offers the electrical conductivity characteristics of the true band-conduction semiconductor. The reason is that in any printed channel made from a semiconducting material, $e.g.$, MoS$_2$, the electrical conductivity will be of the electron hopping type, having little to do with the intrinsic semiconducting properties of the 2D material used to formulate the ink. This situation explains why the electron mobilities in most printed transistors are below 10 cm$^2$V$^{-1}$s$^{-1}$, and in many cases are below 0.5 cm$^2$V$^{-1}$s$^{-1}$.[9,15] Some approaches that have led to mobilities above 10 cm$^2$V$^{-1}$s$^{-1}$ require extra processing steps and utilize nanoparticle inks that are not stable in ordinary solvents.[9] Even in these devices, however, the band-type conduction is not attained.[16] Hence, we face a fundamental problem with achieving semiconducting band-conduction transport in a printed device channel. In any technologically feasible percolated network of semiconducting fillers, electron transport will be limited by the interface between the two contacting fillers. Overcome this challenge that limits the applications of printed electronics requires an unconventional solution.

Recently, charge-density-wave (CDW) materials and devices have attracted renewed interest in the context of 2D van der Waals quantum materials.[17–23] The 1T polymorph of TaS$_2$ is one of the materials from the transition-metal dichalcogenide (TMD) group that reveals several CDW phase transitions detectable *via* resistivity changes and hysteresis.[24,25] A schematic of the crystal structure of the material is presented in Figure 1 (a). The CDW transitions can be induced by temperature and electric bias.[19,21,23] The phase transitions at ~200 K and 350 K are accompanied by a pronounced hysteresis in the current-voltage response. It is worth mentioning that the transition at lower temperatures cannot be observed if the thickness of the material is less than 9 nm. At temperatures below this lower transition (~200 K), the charge density is fully commensurate with the underlying lattice (C-CDW). At this state, the atoms form aggregates of thirteen atoms in the star of David motif illustrated in Figure 1 (b-c). As the temperature increases, the material enters a nearly commensurate phase (NC-CDW) in which the islands of the commensurate phase exist in a sea of incommensurate phase (Figure 1 (d)). Continuing temperature increase drives the material to the incommensurate (IC-CDW) phase at temperatures higher than 350 K (Figure 1 (e)). The hysteresis associated with the transition from the NC-CDW phase to the IC-CDW phase at 350 K has been already exploited in a number of devices with micrometer-scale channels.[23]

In this work, we print two-terminal CDW devices with chemically exfoliated 1T-TaS$_2$ inks on Si/SiO$_2$ substrate. We describe the processing steps required to formulate and prepare inks with 2D CDW filler material, and we demonstrate the operation of the first printed CDW devices. The temperature-dependent electrical transport and low-frequency current fluctuation measurements confirm that the printed devices preserve the CDW transitions, particularly those associated with the C-CDW and NC-CDW phases. Since the functionality of the CDW-based devices is based on the hysteresis and changes in the resistivity of the individual CDW fillers rather than electron hopping conductivity, the operation of such printed devices is not limited by the interfaces between the percolating fillers. This innovative approach, which allows the intrinsic functionality of individual 1T-TaS$_2$ channels to be preserved in large-area printed devices, is a significant advance for printed electronics because it overcomes the obstacle of printing a semiconducting channel for achieving certain functionalities.

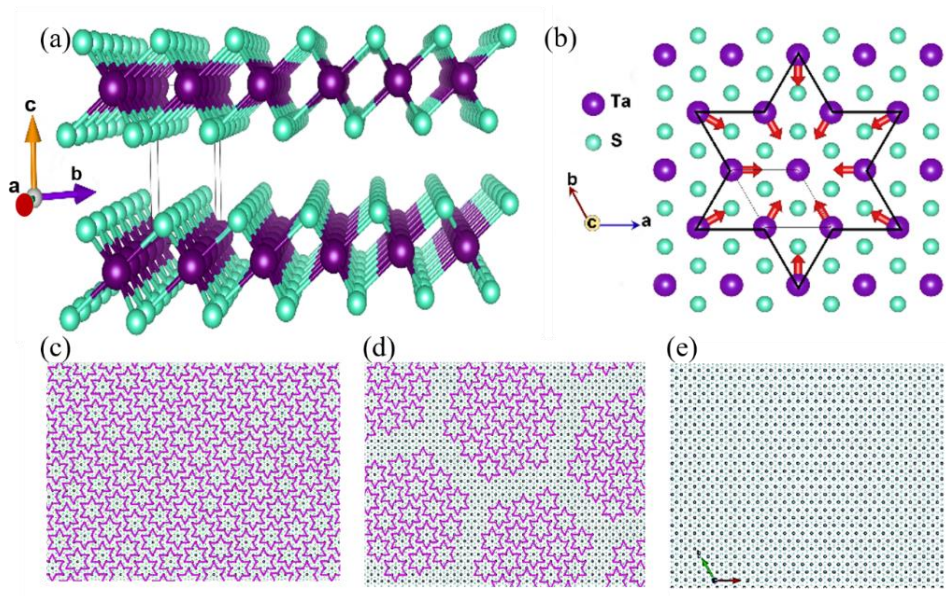

**Figure 1:** (a-b) Crystal structure of 1T-TaS$_2$ along the different axes at the commensurate CDW state. Note the formation of the star-of-David atomic arrangement. (c-e) Schematics of the atoms structuring in (c) fully commensurate, (d) nearly commensurate, and (e) incommensurate CDW phases.

## Results and Discussion

**Ink Preparation and Characterization:** High-quality 1T-TaS$_2$ crystals used for this study were synthesized by the chemical vapor transport (CVT) method using I$_2$ as the transport agent; details have been reported by us elsewhere.[20] The optical and SEM images of the as-grown bulk crystals is shown in Figure 2 (a-b), respectively. We used liquid-phase exfoliation (LPE) to prepare dispersions of few-layer 1T-TaS$_2$ flakes serving as inks. We evaluated several solvents, including acetone, isopropyl alcohol (IPA), dimethylformamide (DMF), and *N*-methylpyrrolidone (NMP), as the liquid medium to exfoliate bulk crystals of 1T-TaS$_2$. Among these solvents, DMF proved to be the most efficient, *i.e.* more crystals were exfoliated in a shorter time. Figure S1 shows a summary of the procedure used for the ink preparation. Figure 2 (c) shows an SEM image of exfoliated 1T-TaS$_2$ in DMF. The lateral dimensions of the fillers were in the range of few microns after LPE. The high boiling point of solvents such as DMF and NMP can lead to lengthy printing processes. For this reason, solvents with lower boiling points, *e.g.*, IPA or ethanol are preferred as the base of many functional inks. The exfoliated fillers of $1T$-$TaS_2$ in DMF were centrifuged and separated from DMF and transferred to 10 mL of IPA as the secondary solvent. The solution was sonicated for 5 minutes to ensure good dispersion of the exfoliated material and eliminate any possible agglomeration of the exfoliated fillers. Figure S2 shows an optical image of cuvettes containing IPA-$TaS_2$ solutions with different concentrations of the fillers, decreasing from left to right. Despite the low boiling point and fast evaporation rate of IPA, these solvents suffer from high surface tension and low viscosity, which often lead to poor ink quality and formation of the coffee-ring effect when printed.[26,27] To mitigate these effects, we used a mixture of two solvents with different boiling points. In this approach, the formation of the coffee rings is suppressed owing to the Marangoni effect in the ink droplet.[28] Specifically, we added the same volumetric amount of ethylene glycol (EG) to the solution of IPA with $1T$-$TaS_2$. The addition of EG to IPA has not only adjusted the boiling point of the ink and eliminated the coffee-ring effect but also tuned the ink's density, viscosity, and surface tension. The adjustment of the thermophysical properties of the ink is crucial for the proper printing process, and it is addressed below. In order to verify the quality of the LPE $1T$-$TaS_2$, we conducted Raman spectroscopy of $1T$-$TaS_2$ before and after the exfoliation at room temperature (RT). The Raman spectra were excited by a blue laser with 488-nm excitation wavelength. The power of the laser was kept at 350 $\mu W$ to minimize the temperature rise, and avoid inducing CDW transition by the laser heating. The results are shown in Figure 2 (d). The blue and purple curves correspond to the Raman spectra of the bulk and exfoliated $1T$-$TaS_2$, respectively. The red curve represents the cumulative fitting on the data for the bulk crystal using individual Gaussian functions. The well resolved Raman peaks were observed at 64, 98, 239, 310, 367, and 390 $cm^{-1}$. The spectral positions of these peaks agree well with the Raman signature of single-crystal $1T$-$TaS_2$, confirming the high quality of the synthesized crystals.[29,30] The spectral positions of the Raman peaks of the exfoliated material match that of the as-synthesized crystal. Broadening of the peaks located at 64 $cm^{-1}$ and 98 $cm^{-1}$ suggests that some defects are introduced to the crystal during the LPE process.

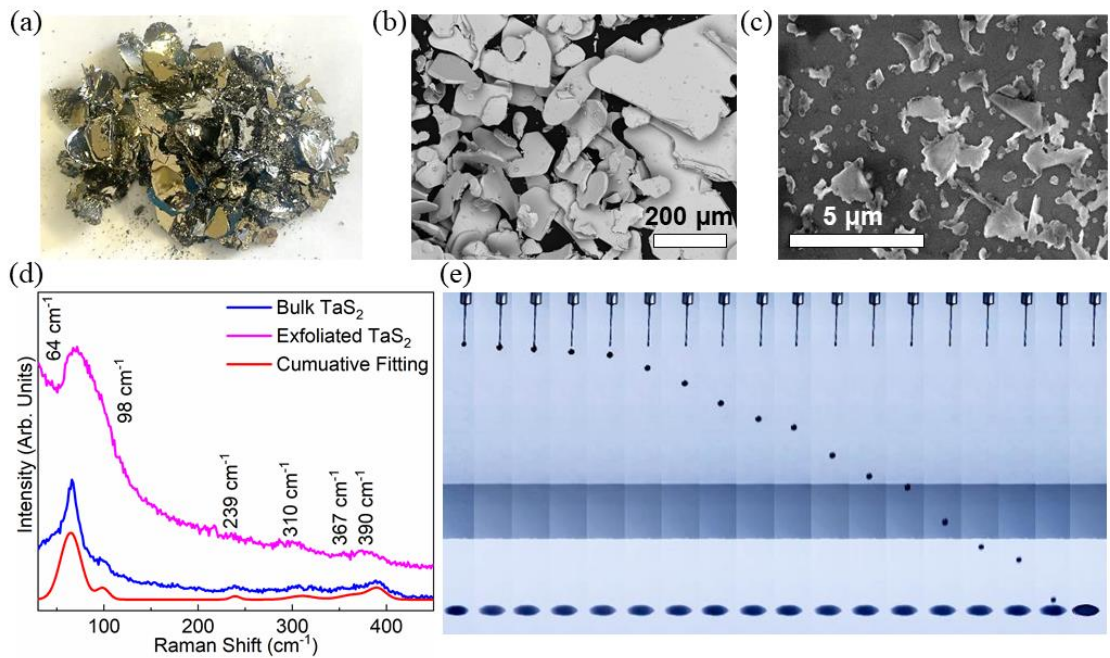

**Figure 2:** (a) Optical image of the as-synthesized bulk 1T-TaS$_2$. (b-c) SEM image of 1T-TaS$_2$ before and after exfoliation in DMF solvent. (d) Raman spectra of 1T-TaS$_2$ before (blue curve) and after (purple curve) liquid phase exfoliation. The red curve is the cumulative fitting of the Raman spectrum shown in blue curve using individual Gaussian functions. (e) Time-lapse optical image of the ink droplet formation, release, and trajectory to the substrate.

Engineering the thermophysical properties of the inks is essential for proper ink droplet formation, dispensing, and spreading of the ink on the surface. The characteristic figure of merit for the ink droplet formation and dispensing is determined by the dimensionless Z-number (or inverse of the Ohnesorge number), $Z = \sqrt{\zeta \rho a}/\mu$, in which $\zeta$ [Nm$^{-1}$], $\rho$ [kgm$^{-3}$] , and $\mu$ [Pa × s] are the ink's surface tension, density, and dynamic viscosity, respectively. The parameter $a$ [m] is the printer's nozzle diameter used in the printing process. Typically, the Z-number should be in the range of $1 \leq Z \leq 14$, otherwise satellite droplet formation ($Z > 14$) or elongated ligaments ($Z < 1$) may occur. These problems can drastically reduce the print quality by dispensing the ink droplets at unwanted areas.[26,31–33] The density, dynamic viscosity, and surface tension of the ink were measured to be $\rho \sim 949.6$ (kgm$^{-3}$), $\mu \sim 5.85 \times 10^{-3}$ (Pa × s), and $\zeta \sim 2.94$ (Nm$^{-1}$), respectively (see Methods for the details of the measurements). We used a modified 3D printer to mimic the functionality of an inkjet printer (see Figure S3 and Methods) with the syringe diameter of $a = 210$ μm. The Z-number based on the ink's thermophysical properties and dispensing nozzle was calculated to be $Z \sim 13.1$. The obtained value is within the defined limits to ensure stable jetting and high print quality. Figure 2 (e), from left to right, shows a time-lapse optical image of the ink droplet formation at the nozzle until it reaches the substrate. We note that no secondary or elongated droplets were observed during the printing process owing to the perfectly tuned Z-number.

**Device Printing:** In this study, several two-terminal devices were fabricated by printing the ink containing chemically exfoliated 1T-TaS$_2$ fillers on top of a set of electrodes prefabricated on a Si/SiO$_2$ substrate. A schematic of the device structure is presented in Figure 3 (a). Each electrode consisted of Ti and Au with the respective thickness of 10 nm and 100 nm on the substrate. The electrodes were fabricated using electron-beam lithography (EBL) and a subsequent lift-off procedure. An optical image of the electrodes before printing the ink is provided in Figure 3 (b). The electrodes are 500 µm long with 3 µm spacing in between. The ink of 1T-TaS$_2$ was printed using a nozzle with an inner diameter of 210 µm. The temperature of the print bed was kept at 70 ℃ to expedite the drying of the ink and further prevent the coffee-ring formation by enhancing the flow of the material through the ink droplet. An optical image of the actual printed device is presented in Figure 3 (c). The dimensions of the printed channel are 3 µm×500 µm×2 µm (L×W×t). The thickness of the channel was measured using an optical profilometer.

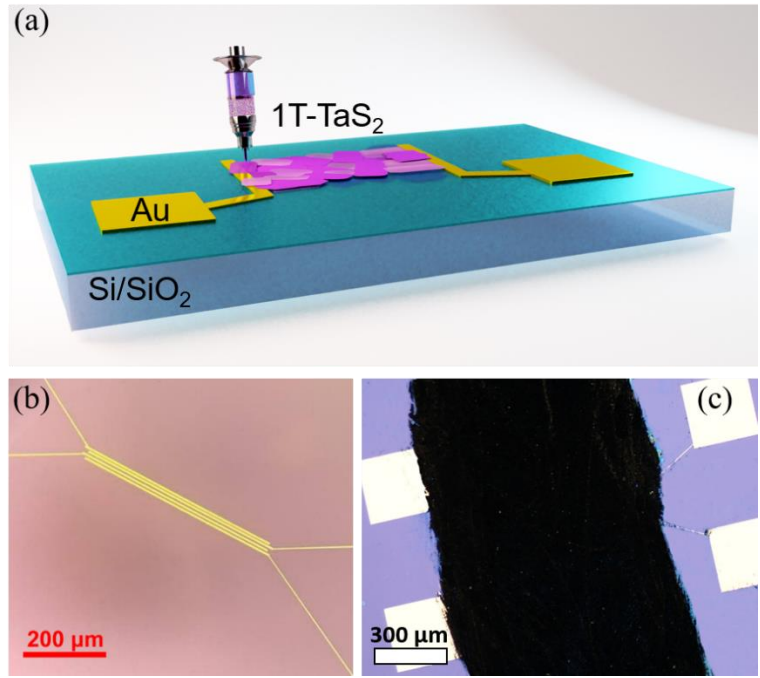

**Figure 3:** (a) Schematic of the device structure and printing process. Optical microscopy images of the (b) fabricated electrodes and (c) actual printed device.

**Device Evaluation:** The current-voltage (I-V) characteristics of the printed devices were measured in the temperature range from 80 K to 400 K. These results are presented in Figure 4 (a). At temperature T > 340 K, the I-V curves are linear suggesting that the contacts of the printed device have Ohmic characteristics. As the temperature decreases, slight deviations from linearity in the I-V curves emerge. The latter is likely due to the transition of 1T-TaS$_2$ fillers from IC-CDW to NC-CDW, and then, to the C-CDW phase with further cooling. The phase transition from IC-CDW to C-CDW is accompanied by a ~0.2 eV band-gap opening in 1T-TaS$_2$.[34,35] Consequently, there is a possibility of Schottky barrier formation at the contact interfaces. The resistance, $R$, of the device at different temperatures for both heating and cooling cycles was extracted from the I-V curves (see Figure 4 (b)). The effective resistivity of the channel at RT, without excluding the contact resistance, is calculated to be $1.6 \times 10^{-5}$ $\Omega$cm$^{-1}$ – substantially larger than that of the single-crystal 1T-TaS$_2$.[36] Possible reasons and implications of this are discussed below.

The most important feature in Figure 4 (b) is the formation of a large hysteresis, attributed to the transition between the NC-CDW to C-CDW phases, or vice versa, during the cooling and heating cycles, respectively. The resistance of the device is increased more than two times during this transition in the cooling cycle. The width of the hysteresis is ~100 K. The change in the resistance of the printed devices at NC-CDW to C-CDW is smaller compared to that of the two-terminal devices fabricated with the single-crystal 1T-TaS$_2$.[21,36] However, the formation of the hysteresis window confirms that the CDW properties of the 1T-TaS$_2$ material are preserved after LPE, mixing with solvents, and printing. The emergence of the NC-CDW to C-CDW phase transition hysteresis indicates that the functionality of the 1T-TaS$_2$ printed devices is defined by the CDW transitions in individual filler nanosheets rather than by electron hopping from nanosheet to nanosheet. The latter suggests that printing can be used to fabricate electronic CDW devices that utilize the hysteresis window for their operation, *e.g.*, CDW-based switches or oscillators.[18] This is an unconventional application of printing with functional inks that contain quasi-2D fillers of van der Waals materials. A small jump in the resistance near T~350 K during the heating cycle possibly can be associated with the NC-CDW to IC-CDW transition. However, it is not observed as clearly as the transition in CDW devices fabricated with single-crystal 1T-TaS$_2$ material.[18,21]

Generally, the dominant mechanism of electron transport in printed devices is electron hopping.[37,38] Electron hopping rather than intrinsic properties of the fillers limit the overall electron mobility and electrical conductivity in channels printed with inks of semiconducting fillers. The observed abrupt change in the resistance and emergence of the hysteresis suggest that in our devices the electrical resistivity of the individual fillers make an important contribution to the overall resistivity of a printed channel. A high background resistance of the device channel is likely due to a combination of the defects in the fillers, contact resistances between them, and effects of the matrix. The switching between C-CDW and NC-CDW phases in individual fillers, induced by either external temperature change or local heating due to passing an electric current, produces sufficient change in the overall resistivity observed in our experiments. Naturally, the measured I-Vs depend on the composition of the ink and CDW filler parameters, and would need to be determined and optimized for each ink and device design.

To further analyze the I-V characteristics in the printed devices, we calculated the differential resistance, dR/dT, of the device for the cooling and heating cycles (Figure 4 (c)). The two sharp peaks observed below the RT are associated with the NC-CDW to C-CDW transitions during the cooling and heating cycles. A small peak in the heating cycle observed at ~350 K is attributed to the NC-CDW to IC-CDW transition as discussed above. Because the hysteresis associated with the NC-CDW to IC-CDW phase transition is substantially smaller than the one associated with the NC-CDW to C-CDW phase transition in single-crystal CDW devices, it is likely that the 1T-TaS₂ nanosheets have partially degraded due to exposure to solvents and atmospheric conditions. For this reason, in this study, we focus on the NC-CDW to C-CDW transition observed at T ~200 K.

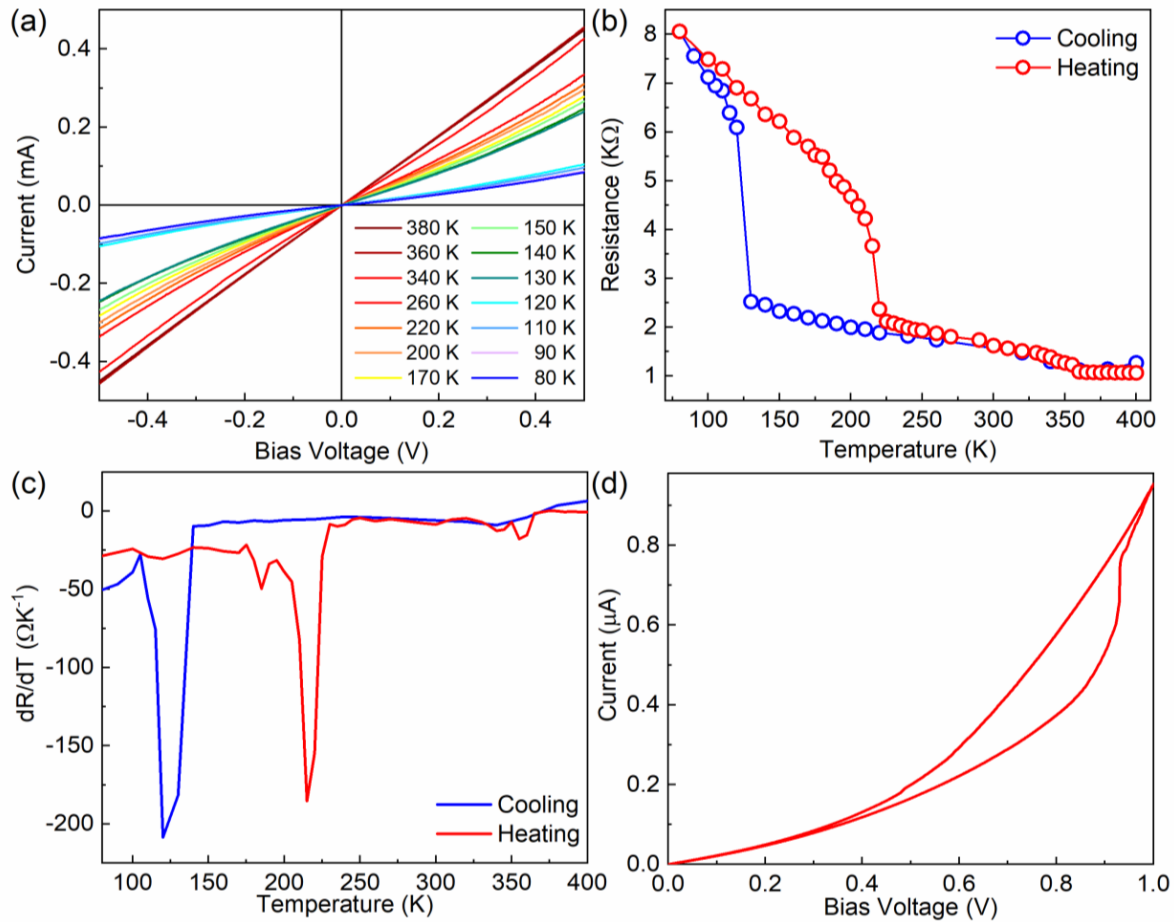

**Figure 4:** (a) Current-voltage curves of the printed device at different temperatures. The device was cooled down from 380 K to 80 K. (b) Resistance, *R*, of the 1T-TaS₂ channel as a function of temperature for cooling (blue) and heating (red) cycles. Note the characteristic hysteresis

window at NC-CDW to C-CDW transition at T~200 K. (c) Differential resistance of the 1T-TaS$_2$ channel as a function of temperature for the cooling and heating cycles. The dominant peaks below the room temperature are associated with the NC-CDW to C-CDW transition. The small peak in the red curve at T~350 K is attributed to the NC-CDW to I-CDW transition. (d) Current-voltage characteristic of the device at 110 K as a function of the bias voltage.

The CDW transition in 1T-TaS$_2$ can be induced by stimuli other than the external temperature change.[19] For example, application of a bias voltage may also trigger the CDW transition. Depending on the device size and design, the CDW phase transition can be induced either by the electric field directly or by Joule heating as a result of applying electric bias and passing the current.[19,23] Figure 4 (d) shows the I-V characterization of the printed device at $T = 110$ K, below the C-CDW to NC-CDW phase transition temperature. When the bias exceeds ~0.9 V, an abrupt jump in the I-V curve is observed. The change in the resistance is attributed to the C-CDW to NC-CDW phase transition. We note that a full hysteresis, similar to the one shown in Figure 4 (b), was not observed in the I-V measurements. A full hysteresis induced by passing an electric current is routinely observed in devices fabricated with individual 1T-TaS$_2$ layers.[19,23,39] The fact that we have not observed it in the printed devices is likely related to the inherent disorder of the fillers in the ink, non-uniform heating and large size of the devices.

We measured the low-frequency current fluctuations, *i.e.* excess electronic noise, to further elucidate the phase transitions in the printed device[20,23] Low-frequency noise measurements have been used to understand electronic transport and assess the reliability of the devices.[40–45] The details of the experimental setup and measurement procedures have been reported by some of us elsewhere.[39,44] At frequencies below 100 kHz, electrically conductive materials often reveal current fluctuations with the spectral density $S(f) \sim 1/f^\gamma$, where $f$ is the frequency and parameter $\gamma \sim 1$. Figures 5 (a-b) show the voltage-referred noise power spectral density, $S_v$, and the normalized current noise power spectral density, $S_I/I^2$, as a function of frequency at different temperatures. In all experiments, the bias voltage was kept at ~0.1 V to avoid Joule heating. One can see that the noise spectrum follows the $1/f$ trend. Some traces of the Lorentzian bulges at frequencies above $f$=100 Hz are observed that might be indicative of defects or impurities.[142] In Figure 5 (c-d), we present the noise power spectral density and the normalized current noise power spectral density at the frequency of $f$=10 Hz as a function of temperature. Noise measurements were carried out when the device was cooled down from 400 K to 80 K with a cooling rate of 2 K/min. The C-CDW to NC-CDW phase transition at T ~200 K is clearly observed as an abrupt increase in the spectra, whereas the NC-CDW to IC-CDW at T~350 K is also visible but less pronounced (insets in Figure 4 (c-d)). The temperatures at which the phase transitions happen agree well with the resistance data shown in Figure 4 (b-c).

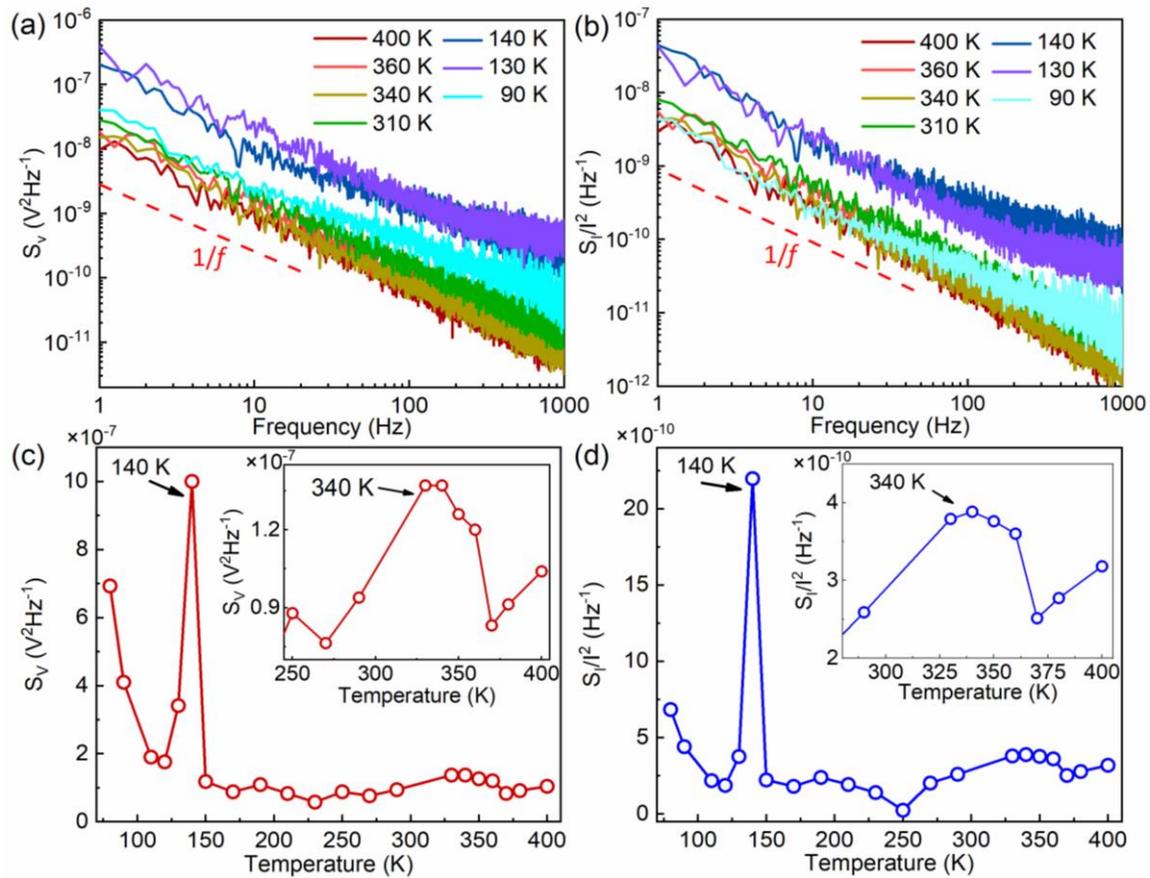

**Figure 5:** Temperature-dependent low-frequency electronic noise characteristics of the printed devices. (a) Low-frequency noise spectra of voltage fluctuations, $S_v$, as a function of frequency measured at different temperatures. as a function of frequency at different temperatures. The device has been cooled down to 80 K. (b) Normalized current noise spectral density at different temperatures. (c-d) $S_v$ and normalized current noise spectral density as a function of temperature at the constant frequency of $f = 10$ Hz. The device voltage was kept constant at 0.1 V.

The focus of the present work was to demonstrate the possibility of printing with an ink comprised of a quasi-2D CDW material, specifically 1T-TaS$_2$, and to verify that standard ink processing preserves functional CDW phases. Investigation of the switching mechanism and electronic transport in the printed device channels requires specially designed experiments with a small number of overlapping 1T-TaS$_2$ nanosheets in order to disentangle the electronic hopping from the phase transitions and collective currents in the individual nanosheets. These tasks are clearly beyond the scope of the present study. At this point, we can make a few preliminary conclusions about the functionality and operation of these printed CDW devices. The broadening of some of the characteristic Raman peaks, e.g. at 64 cm$^{-1}$ and 98 cm$^{-1}$ (see Figure 2 (b)) suggests the emergence of defects in 1T-TaS$_2$ fillers introduced during the LPE and printing processes. The substantially higher effective resistivity of the printed channels compared to individual 1T-TaS$_2$ flakes indicates several possible reasons, including the role of the contact resistance with the metal electrodes, contact resistance between overlapping flakes, defects within individual 1T-TaS$_2$ flakes, and the electron hopping transport mechanism. All these factors together can contribute to the observed two orders of magnitude increase in the effective resistance of the channel.

To explain the observed change in the resistance and appearance of the hysteresis at the CDW transition temperature of ~200 K, one has to assume that for a certain combination of 1T-TaS$_2$ filler dimensions and ink properties, the resistance change due to the CDW phase transition in the individual fillers is comparable to the overall resistance of the printed channel. We have previously shown that CDW transitions in individual layers of 1T-TaS$_2$, with ~10-nm-scale thickness and ~5-μm-scale lateral dimensions placed on thermally resistive substrates are due to local heating resulting from passing the current.[19,41] Due to the large size of the printed devices and low current levels, the channel heating is estimated to be too small to induce CDW phase transitions in all 1T-TaS$_2$ fillers. One has to assume a non-uniform voltage drop, *e.g.* happening mostly over the fillers in the proximity to the metal electrodes, to produce sufficient local heating to drive the CDW phase transition in a subset of CDW filler nanosheets, while others are just contributing to the overall series resistance.

The field of CDW materials and devices is evolving toward engineering domains with potential applications for amplifiers, detectors, memory and optoelectronic devices, as well as information processing and radiation-hard electronics.[17,20,46–50] This demonstration of printed CDW devices with hysteresis and basic resistive switching by an applied voltage is an important step for the practical implementation of CDW materials. Many demonstrated $1T-TaS_2$ CDW devices utilize the hysteresis in I-V curves for their functionality. The incorporation of printing technologies to achieve CDW devices on flexible substrates increases their application domain substantially.

**Conclusions**

Printed electronics face a fundamental problem – one cannot achieve the semiconducting band-conduction transport in a printed device. A printed channel with the ink of semiconducting fillers never offers an electrical conductivity characteristic of the true band-conduction semiconductor. In any technologically feasible percolated network of semiconducting fillers, the electron transport is limited by the interface between the two contacting fillers. The latter limits the functionality of printed electronics. We addressed this problem by developing techniques for the formulation of inks containing fillers derived from the quasi-2D CDW material, $1T-TaS_2$, and inkjet printing of CDW electronic devices. The temperature-dependent electrical and current fluctuation measurements demonstrated that the CDW properties of $1T-TaS_2$ fillers were preserved in the printed thin films. The functionality of the printed thin-film devices was defined by the nearly-commensurate to commensurate CDW phase transition in the individual $1T-TaS_2$ fillers rather than by electron-hopping transport between the fillers. These results provide pathways for the development of printed electronics with much greater functionality achieved via the use of quasi-2D van der Waals quantum materials.

## METHODS

**1T-TaS₂ Ink Preparation and Characterization:** To obtain 1T-TaS$_2$ 2D fillers, 3 mg of the bulk crystals were placed in a vial containing 10 mL of DMF. The mixture was sonicated in a bath sonicator for 90 min. Then any unexfoliated crystals were separated from the dispersion and sonicated in 5 mL of fresh DMF for 90 min. This process was repeated until no unexfoliated crystals remained. The combined supernatant dispersion was centrifuged at 11000 rpm for 15 min. This led to the separation of exfoliated 1T-TaS$_2$ from the DMF. Then the DMF was removed, and the fillers were dispersed in 10 mL of IPA as the secondary solvent. The dispersion was sonicated for 5 minutes to ensure homogeneity and reduce agglomeration. This step was repeated three times to remove residual DMF. The purpose of adding EG was to increase the viscosity of the ink and ensure proper dispensing and drying of the ink. The concentration of exfoliated fillers in the ink was measured with an Agilent 60 UV/Vis spectrophotometer. For measuring the viscosity of the ink, a Cannon SimpleVIS viscometer was used with 0.5 μL of the ink inserted in the device. The measured kinematic viscosity displayed on the device was then divided by density to calculate the dynamic viscosity. The surface tension of the ink was measured utilizing a CSC Scientific DuNouy interfacial tensiometer. This was done by inserting the device ring into the ink and raising the ring slowly until the film between the ink and the ring breaks. The measured amount of surface tension can then be read from the from the device dial. After the addition of EG, the viscosity and the surface tension of the ink were measured to be $5.85 \times 10^{-3}$ (Pa × s) and $\gamma \sim 2.94$ $(N/m)$. The density of the ink was calculated using the rule of mixtures to be 949.6 (kg/m$^3$). The calculated Z number from the measured ink properties is 13.1 considering the inner diameter of the nozzle (210 μm) used for this work.

**Printing Devices with 1T-TaS₂ CDW Materials:** The 1T-TaS$_2$ ink was used in a modified 3D printer (Hyrel 30M) in which the printer head was swapped with a syringe holder to mimic the functionalities of an inkjet printer. A blunt-end needle syringe with gauge 27 was used for the dispensing of the ink. During the printing process, the bed temperature of the printer was held at 70 °C. 10 layers of material were printed on an Si/SiO$_2$ substrate with gold electrodes. The electrodes were fabricated using electron-beam lithography, metallization of Ti/Au (10-/100-nm) by an e-beam evaporator and finally lift-off of excess metal. The final printed 1T-TaS$_2$ device had the channel length, width, and thickness of 3 µm, 500 µm, and 2 µm, respectively. The thickness of the printer channel was measured with a surface profilometer (Profilm 3D).

**Electrical Measurement and Noise Spectroscopy:** The current–voltage (I–V) characteristics were measured in the cryogenic probe station (Lakeshore TTPX) with a semiconductor analyzer (Agilent B1500). The two terminal device configuration was used for the low-frequency noise experiments. The noise spectra were measured with a dynamic signal analyzer (Stanford Research 785). The bias voltage was supplied by a battery biasing circuit to minimize the noise at 60 Hz and the harmonics associated with it. The signal measured by the dynamic signal analyzer is the absolute voltage noise spectral density, S$_v$, of a parallel resistance network consisting of a load resistor (R$_L$) of 3 KΩ and the device under test with a resistance of R$_D$. The normalized current noise spectral density, S$_I$/I$^2$, was calculated $S_I/I^2 = S_V \times [(R_L + R_D)/(R_L \times R_D)]^2/(I^2 \times G)$ where G is the amplification of the low-noise amplifier.

## Acknowledgements

The work at UC Riverside was supported, in part, by the U.S. Department of Energy Office of Basic Energy Sciences under the contract No. DE-SC0021020 "Physical Mechanisms and Electric-Bias Control of Phase Transitions in Quasi-2D Charge-Density-Wave Quantum Materials".

## Author Contributions

A.A.B. conceived the idea, coordinated the project, and contributed to experimental data analysis; F.K. contributed to data analysis and led the manuscript preparation; S.B. and Z.B. conducted chemical exfoliation, device printing and electrical testing; A.M. assisted with the lithography and electronic noise measurements; Y.G. synthesized bulk crystals of 1T-TaS2; T.T.S. supervised material synthesis and contributed to data analysis; S.S. assisted with chemical exfoliation and ink preparation. All authors contributed to the manuscript preparation.

**Supplemental Information**

The supplemental information is available at the journal web-site for free of charge.

**The Data Availability Statement**

The data that support the findings of this study are available from the corresponding author upon reasonable request.